\def\o{\over}
\def\A{\rightarrow}
\def\bar{\overline}
\def\kp{K\pi}
\def\pp{\pi\pi}
\def\r{\gamma}
\def\a{\alpha}
\def\b{\beta}
\def\p{\pi}

\def\Vtb{$V_{tb}$}

\def\ba{b_{L\alpha}}
\def\bb{b_{L\beta}}
\def\sbb{{\bar s}_{L\beta}}
\def\sba{{\bar s}_{L\alpha}}

\def\bar{\overline}

\def\G{{\rm GeV}}
\documentstyle [12pt]{article}
\begin{document}
\baselineskip=25pt
\setcounter{page}{1}
\thispagestyle{empty}
\begin{flushright}
\begin{tabular}{c c}
& {\normalsize   AUE-05-93}\\
& {\normalsize  EHU-06-93}\\
& {\normalsize  KGU-03-93}\\
& {\normalsize  October 1993}
\end{tabular}
\end{flushright}
\vspace{0.5cm}
\centerline{\Large\bf Penguin Effects on $\kp$ and $\pp$ Decays
 of the $B$ Meson}
\vskip 1.2 cm
\centerline{{\bf Takemi HAYASHI}$^{(a)}$, {\bf Masahisa MATSUDA}$^{(b)}$
 \footnote{E-mail:masa@auephyas.aichi-edu.ac.jp}
and {\bf Morimitsu TANIMOTO}$^{(c)}$}
\vskip 0.5 cm
\centerline{$^{(a)}$ \it{Kogakkan University, Ise, Mie 516, JAPAN}}
\centerline{$^{(b)}$ \it{Department of Physics and Astronomy, Aichi University
of Education}}
\centerline{\it Kariya, Aichi 448, JAPAN}
\centerline{$^{(c)}$ \it{Science Education Laboratory, Ehime University}}
\centerline{\it Matsuyama, Ehime 790, JAPAN}
\vskip 1.2 cm
\centerline{\bf ABSTRACT}
\vskip 0.15 cm
   We give the detailed analyses for the gluonic-penguin effect
on the $\kp$ and $\pp$ decays  of the $B$ meson. In the standard model,
it is shown that the ratio $BR(B\A\kp)/BR(B\A\pp)$ takes the value
$0.5\sim 3.0$ with the strongly depending on the CP violating phase $\phi$
and the KM matrix element $|V_{ub}|$.
We obtain the constraint on the form factor by using the experimental
branching ratio.
It is also found that, in the two-Higgs-doublet model,
 the charged Higgs contribution which
could  enhance the  $B\A X_s\r$ decay does not a sizable effect
on our processes.
The effect of the final state interaction on these processes is also discussed.
\newpage
\topskip 0.05 cm
 Until now,  the rare $B$ decays have been intensively studied  in the
standpoint of the standard model(SM) and also beyond the standard model.
 Especially, $b\A s\r$ and $b\A s g$ sub-processes have attracted one's
attention  in the circumstance that
the experimental evidence has been found in CLEO
\cite{BSR}.
These decays, induced by the flavour changing neutral current, are
controlled by the one-loop penguin operators
which involve the important SM parameters
such as the top-quark mass and the Kobayashi-Maskawa matrix elements
$V_{ts}$ and \Vtb\cite{KM}.
In our previous papers
\cite{HAYASHI},
 we analyzed the inclusive decay $B\A X_s\r$ and the exclusive decays
$B\A K_X\r$,
where $K_X$ denotes the meson states in the
$s\bar q$($\bar q=\bar u\ {\rm or}\ \bar d$) system,
     as well as the exclusive decays
 $B\A K_X\phi$ by including the nonstandard physical effects due to
the charged Higgs contributions in the two-Higgs-doublet model(THDM)
\cite{DAVIES}\cite{NEWHAYASHI}.\par
 In this paper, we study the $B\A \kp$ and  $B\A \pp$ decays,  which are
induced by both tree processes and  gluonic penguin ones, in the SM model.
 Now we take into account QCD corrections, which were not included
in the previous calculations\cite{OLD}, since the QCD corrections
 have been  found to give the important contributions to the rare decays of the
 $B$ meson.
Recently, CLEO Collaboration reported the following experimental result
\cite{CLEO}:
\begin{equation}
BR(B^0_d\A \pi^+\pi^-+ K^+\pi^-)=(2.4^{+0.8}_{-0.7}\pm 0.2)\times 10^{-5}
\ ,
\end{equation}
 \noindent
where the discrimination between the $K^+$ meson and the $\p^+$ meson
has not been suceeded.
We will predict the ratio of the  $B\A \kp$ decay width to the  $B\A \pp$ one,
 the ratio being almost independent of the form factor.
The magnitude of the relevant form factor can be restricted by
the experimental branching ratio of eq.(1) as shown later. Furtheremore, we
 will examine the nonstandard effects due to the charged Higgs contribution
 in the THDM. Finally,
 the effect of the  phase shifts  due to the final state interactions
 will be discussed as to our numerical results.\par
%
  In the previous paper\cite{NEWHAYASHI},
 we give the formulation including the QCD corrections for the
rare decays of the $B$ meson.
 The four-quark operators
and the magnetic-transition-type ones are given for the
processes under consideration in the the Hamiltonian
\cite{GRIN}\cite{BUCHALLA},
\begin{equation}
H_{eff}={4G_F \over \sqrt{2}}\left [v_u\sum_{i=1}^2 C_i(\mu)O_i(\mu)+
v_t\sum_{i=3}^8 C_i(\mu)O_i(\mu)\right ] \ ,
\end{equation}
\noindent where
the  factor  $v_q$(q=u,t) is defined by
\begin{equation}
 v_q= \left\{\matrix{V_{qb}V_{qs}^* \quad {\rm for}\ b\A s \cr
   V_{qb}V_{qd}^*\quad {\rm for}\ b\A d}\right.
\ .
\end{equation}
For the $b\A s$ transition, each operator $O_i$ is  defined as follows:
\begin{eqnarray}
           O_1 &=&({\bar q}_{L \alpha}\gamma^\mu \bb)(\sbb \gamma_\mu
q_{L\alpha})\ , \nonumber\\
           O_2 &=&({\bar q}_{L\a}\gamma^\mu \ba)(\sbb \gamma_\mu q_{L\beta}) \
,\nonumber\\
           O_3 &=&(\sba\gamma^\mu \ba)(\sum_{\rm 5\; quarks}{\bar q'}_{L\beta}
            \gamma_\mu q'_{L\beta}) \ ,\nonumber\\
           O_4 &=&(\sba\gamma^\mu \bb)(\sum_{\rm 5\; quarks}{\bar q'}_{L\beta}
            \gamma_\mu q'_{L\alpha}) \ ,\\
           O_5 &=&(\sba\gamma^\mu \ba)(\sum_{\rm 5\; quarks}{\bar q'}_{R\beta}
            \gamma_\mu q'_{R\beta}) \ ,\nonumber\\
           O_6 &=&(\sba\gamma^\mu \bb)(\sum_{\rm 5\; quarks}{\bar q'}_{R\beta}
            \gamma_\mu q'_{R\alpha}) \ ,\nonumber\\
           O_7 &=&-i{e \over {8\pi^2}}m_b\sba \sigma^{\mu\nu}b_{R\alpha}q_\mu
           \epsilon_\nu \ ,\nonumber\\
           O_8 &=&-i{g_c \over {8\pi^2}}m_b\sba
\sigma^{\mu\nu}T^a_{\alpha\beta}           b_{R\beta}q_\mu \epsilon^a_\nu \ .
\nonumber
\end{eqnarray}
\noindent
For the $b\A d$ transition, the $s$-quark is replaced by the $d$-quark
in each of eq.(4).
  In addition to these operators, we need a new operator $O'_8$, which is
derived from $O_8$ through the coupling of the virtual gluon to $\bar {q'} q'$:
\begin{equation}
           O_8' =im_b{\alpha_s \over 2\pi}{1 \over q^2}
           \bar q_{L\a} \sigma^{\mu\nu}T^a_{\alpha\beta}
           b_{R\beta}q_\mu {\bar q'}_\beta \gamma_\nu
           T^a_{\beta\alpha}q'_\alpha  \qquad (q=s\ {\rm or}\ d) \ ,
\end{equation}
where  $q_\mu$ denotes  the four-momentum of the virtual gluons.
\noindent
The coefficients relevant to the processes $B\A\kp$ and $\pp$ are defined
at the energy scale of $m_W$ as\cite{GRIN}\cite{BUCHALLA}
\begin{eqnarray}
C_1(m_W)&=&0 \ ,\hskip 2.8 cm  C_2(m_W)=1 \ , \nonumber\\
C_3(m_W)&=&C_5(m_W)+{\alpha \over{6\pi\sin\theta^2_W}}\times \nonumber \\
    & &[{1\o 2} ({x_t \over 1-x_t}+{x_t \ln x_t\over (1-x_t)^2})
+{x_t\o 8}({x_t-6 \o x_t-1}+{3x_t+2\o (1-x_t)^2}\ln{x_t})] \ , \nonumber\\
C_5(m_W)&=&-{\alpha_s(m_W) \o 288\pi}G_1(x_t) \ ,\\
C_4(m_W)&=&C_6(m_W)={\alpha_s(m_W) \o 96\pi}G_1(x_t) \ ,\nonumber\\
C_7(m_W)&=&F(x_t) \ ,\hskip 2 cm C_8(m_W)=-{1 \o 8}G_2(x_t)\ . \nonumber
\end{eqnarray}
The functions $G_i(x_t)$ and $F(x_t)$ are given by
\begin{eqnarray}
 G_1(x_t)&=& {{x_t(1-x_t)(18-11x_t -x_t^2)-2(4-16x_t+9x_t^2)\ln x_t}\over
          (1-x_t)^4}\ ,  \nonumber\\
 G_2(x_t)&=&x_t{{(1-x_t)(2+5x_t-x_t^2)+6x_t\ln x_t}\over (1-x_t)^4}\ ,
 \nonumber \\
F(x_t)&=&{x_t \over {24(1-x_t)^3}}
 [8x_t^2+5x_t-7+{6x_t(3x_t-2) \o (1-x_t)}\ln{x_t}] \ ,
\end{eqnarray}
with $x_t=m_t^2/m_W^2$.
%
\par
In the following analysis, we take the value of the coefficient $C_8'(\mu)$
 being equal to $C_8(\mu)$.
Although $C_8(\mu)$ does not include the full QCD correction of $C_8'(\mu)$
 in the leading {\it log} approximation, this replacement does not seriously
affect the results numerically since the $O_8'$ term is the next leading one
compared to the operators $O_3, O_4, O_5$ and $O_6$.
The similar operator $O'_7$, which is induced from $O_7$ by the coupling of
the virtual photon with $\bar q' q'$, is negligible due to $\a\ll \a_s$.
%
We evolve the coefficients $C_i(\mu)$, by starting from the scale $m_W$
  as given in eq.(6) to
the scale $\mu=m_b=4.58$GeV, according to the renormalization group equation
\cite{MASA}.
Then, we obtain
\begin{eqnarray}
C_1(m_b)&=&-0.240 \ ,\hskip 2.8 cm   C_2(m_b)=1.103 \ ,\nonumber\\
C_3(m_b)&=&0.011+1.125C_3(m_W)-0.121C_4(m_W) \ ,\nonumber\\
C_4(m_b)&=&-0.025-0.291C_3(m_W)+0.824C_4(m_W) \ ,\nonumber\\
C_5(m_b)&=&0.007+0.944C_3(m_W)+0.083C_4(m_W) \ ,\\
C_6(m_b)&=&-0.030+0.229C_3(m_W)+1.465C_4(m_W) \ ,\nonumber\\
C_7(m_b)&=&-0.199+0.629C_3(m_W)+0.931C_4(m_W)+0.675C_7(m_W) \nonumber\\
        &+&0.091C_8(m_W) \ ,\nonumber\\
C_8(m_b)&=&-0.096-0.598C_3(m_W)+1.029C_4(m_W)+0.709C_8(m_W)\ . \nonumber
\end{eqnarray}
%
However, the coefficients $C_3(m_b)$, $C_4(m_b)$, $C_5(m_b)$ and $C_6(m_b)$
do not  completely involve the charm-quark loop effect.
The charm-quark loop contributions are included by replacing
these coefficients as follows\cite{FLEI}:
\begin{eqnarray}
C_3(\mu) &\A& C_3(\mu)+{\a_s(\mu)\o 24\pi}G(m_c,q,\mu)C_2(\mu) \ ,\nonumber\\
C_4(\mu) &\A& C_4(\mu)-{\a_s(\mu)\o 8\pi}G(m_c,q,\mu)C_2(\mu)\ ,\nonumber\\
C_5(\mu) &\A& C_5(\mu)+{\a_s(\mu)\o 24\pi}G(m_c,q,\mu)C_2(\mu)\ ,\nonumber\\
C_6(\mu) &\A& C_6(\mu)-{\a_s(\mu)\o 8\pi}G(m_c,q,\mu)C_2(\mu)\ ,\nonumber\\
G(m_c,q,\mu)&=&-4\int_0^1 x(1-x)\ln\left[{m_c^2-q^2x(1-x)\o\mu^2}\right ]dx
\ ,
\end{eqnarray}
\noindent
where the parameter $q^2$ denotes the square of the four-momentum of the
virtual gluons.
These values of the coefficients are numerically given as follows:
\begin{eqnarray}
C_1(m_b) &=& -0.240\ , \hskip 2.8 cm C_2(m_b)= 1.103 \ ,\nonumber\\
C_3(m_b) &=& 0.0152- 0.0058i \ ,\qquad C_4(m_b)= -0.0380+ 0.0174i \ ,
\nonumber\\
C_5(m_b) &=& 0.0118-  0.0058i\ ,\qquad
C_6(m_b)= -0.0427 + 0.0174i \ ,\nonumber\\
C_7(m_b) &=& -0.320 \ ,\hskip 2.8 cm
C_8(m_b) = -0.157 \ ,
\end{eqnarray}
\noindent
where $q^2$ is taken as $m_b^2/2$
\cite{NEWHAYASHI}\cite{FLEI}.
The imaginary part of these coefficients follows from the loop integral
$G(m_c,q,\mu)$.
These values are in agreement with the ones given by Fleischer\cite{FLEI}.
\par
Let us begin with showing the decay amplitude of the
$B\A\kp$ process.
By the use of the above operators, the decay amplitude is written as
\begin{eqnarray}
\langle K\p \mid H_{eff}\mid B\rangle
 & =& {4 G_F\over \sqrt{2}} V_{ub} V^*_{us} \sum_{1,2}C_i(\mu)
 \langle K\p \mid O_i(\mu) \mid B\rangle  \nonumber \\
 &+&{4 G_F\over \sqrt{2}}V_{tb} V^*_{ts} \sum_{3,4,5,6,8'}C_i(\mu)
 \langle K\p \mid O_i(\mu) \mid B\rangle \ .
\end{eqnarray}
\noindent
We use the factorization approximation
in order to estimate  the hadronic matrix element.
This factorization assumption  successfully works
 in the $D$ meson  and $B$ meson decays
\cite{FACTORIZATION}
 within the factor two as to the branching ratios.
Under this assumption,
 the hadronic matrix elements of the above operators are given as
\begin{eqnarray}
 \langle K^+\p^-\mid O_1\mid B^0_d\rangle&=&
- {1\over 12}\langle K^+\mid \bar s\r_\mu\r_5 u\mid 0 \rangle
\langle \p^-\mid \bar u\r^\mu  b\mid B^0_d\rangle\ , \nonumber\\
 \langle K^+\p^-\mid O_2\mid B^0_d\rangle&=&
-{1\over 4} \langle K^+\mid \bar s\r_\mu\r_5 u\mid 0 \rangle
\langle \p^-\mid \bar u\r^\mu  b\mid B^0_d\rangle\ , \nonumber\\
 \langle K^+\p^-\mid O_3\mid B^0_d\rangle&=&
- {1\over 12}\langle K^+\mid \bar s\r_\mu\r_5 u\mid 0 \rangle
\langle \p^-\mid \bar u\r^\mu  b\mid B^0_d\rangle\ , \nonumber\\
 \langle K^+\p^-\mid O_4\mid B^0_d\rangle&=&
- {1\over 4}\langle K^+\mid \bar s\r_\mu\r_5 u\mid 0\rangle
\langle \p^-\mid \bar u\r^\mu  b\mid B^0_d\rangle\ , \nonumber\\
 \langle K^+\p^-\mid O_5\mid B^0_d\rangle&=&
- {2\over 3}\langle K^+\mid \bar s_L u_R\mid 0\rangle
\langle \p^-\mid \bar u_R  b_L\mid B^0_d\rangle\ , \nonumber\\
 \langle K^+\p^-\mid O_6\mid B^0_d\rangle&=&
-2\langle K^+\mid \bar s_L u_R\mid 0\rangle
\langle \p^-\mid \bar u_R  b_L\mid B^0_d\rangle\ , \nonumber\\
 \langle K^+\p^-\mid O_8'\mid B^0_d\rangle&=&
-{\alpha_s\over 12 \pi}m_b{1\o q^2}q^\mu\left(
\langle K^+\mid \bar s\r_\mu\r_5 u\mid 0\rangle
\langle \p^-\mid \bar u  b\mid B^0_d\rangle \right.\nonumber\\
&+& \left.\langle K^+\mid \bar sr_5 u\mid 0\rangle
\langle \p^-\mid \bar u\r_\mu  b\mid B^0_d\rangle\right ) \ .
\end{eqnarray}
 \noindent
By the use of these equations,
the decay amplitude of $B\A \kp$  is given as follows:
\begin{eqnarray}
 \langle  K^+\pi^-\mid H_{eff}\mid B^0_d\rangle &=&
{G_F\o \sqrt{2}}\left [-V_{ub}V_{cs}^*\left ({1\o 3}C_1 +C_2 \right )
a^{\kp}(Q^2) - V_{tb}V_{ts}^* \left ({1\o 3}C_3 a^{\kp}(Q^2)\right.\right.
\nonumber \\
 + C_4 a^{\kp}(Q^2)& +& {2\o 3}C_5 b^{\kp}(Q^2)
+2 C_6 b^{\kp}(Q^2)+{\a_s\o 3\pi}C_8{m_b\o q^2} c^{\kp}(Q^2)\left.\right )
\left.\right ]\ ,
\end{eqnarray}
\noindent where
\begin{eqnarray}
a^{\kp}(Q^2)&\equiv& \langle K^+\mid \bar s\r_\mu\r_5 u\mid 0\rangle
 \langle \pi^-\mid \bar u\r^\mu b\mid B^0_d\rangle \ , \nonumber\\
b^{\kp}(Q^2)&\equiv& \langle K^+\mid \bar s\r_5 u\mid 0\rangle
 \langle \pi^-\mid \bar u b\mid B^0_d\rangle\ , \\
c^{\kp}(Q^2)&\equiv&q^\mu\left(\langle K^+\mid \bar s\r_\mu\r_5 u\mid 0\rangle
 \langle \pi^-\mid \bar u b\mid B^0_d\rangle+
 \langle K^+\mid \bar s\r_5 u\mid 0\rangle
 \langle \pi^-\mid \bar u \r_\mu b\mid B^0_d\rangle\right )\ . \nonumber
\end{eqnarray}
\noindent
The  hadronic matrix elements $a^{\kp}$, $b^{\kp}$ and $c^{\kp}$ are
given in terms of the decay constant $f_K=161$MeV and
the longitudinal  form factor $F_0^{B\pi}(Q^2)$ as
\begin{eqnarray}
a^{\kp}(Q^2)&=&f_K (m_B^2-m_{\pi}^2)F_0^{B\pi}(Q^2)\ , \nonumber\\
b^{\kp}(Q^2)&=&{m_K^2\o (m_s+m_u)(m_b-m_u)}a^{\kp}(Q^2)\ , \nonumber\\
c^{\kp}(Q^2)&=&\left [{{m_B^2-m_K^2\o 2(m_b-m_u)}+{m_K^2\o m_s+m_u}{m_B^2\o
m_B^2-m_\pi^2}} \right ]a^{\kp}(Q^2) \ ,
\end{eqnarray}
\noindent where $f_K$ and the relevant form factors are defined by
\begin{eqnarray}
 \langle K^+\mid \bar s\r_\mu\r_5 u\mid 0\rangle &=&f_K Q_\mu \ , \\
 \langle \pi^-\mid \bar u\r_\mu b\mid B^0_d\rangle &=& \left (p_B+p_\pi-
{m^2_B-m^2_\pi\o Q^2}Q\right )_\mu F_1^{B\p}(Q^2)+
 {m^2_B-m^2_\pi\o Q^2}Q_\mu F_0^{B\p}(Q^2) \ , \nonumber
\end{eqnarray}
\noindent with $Q=p_B-p_\pi$ and then with $Q^2=m_K^2$.
In the estimation of $b^{\kp}$ and $c^{\kp}$,
the equations of motion are used for quarks and anti-quarks.
Then, the decay branching ratio is obtained by calculating
\begin{equation}
 BR(B_d^0\A K^+\pi^-)=\tau_B{p_\pi\o 8\pi m_B^2}
      \mid \langle  K^+\pi^-\mid H_{eff}\mid B_d^0\rangle \mid^2 \ .
\end{equation}
\par
 The decay amplitude of  $B\A\pp$ is given in the same way,
\begin{eqnarray}
 \langle  \pi^+\pi^-\mid H_{eff}\mid B^0_d\rangle &=&
{G_F\o \sqrt{2}}\left [-V_{ub}V_{cd}^*\left ({1\o 3}C_1 +C_2 \right )
a^{\pp}(Q^2) - V_{tb}V_{td}^* \left ({1\o 3}C_3 a^{\pp}(Q^2)\right.\right.
\nonumber\\
 + C_4 a^{\pp}(Q^2) &+& {2\o 3}C_5
b^{\pp}(Q^2)  +2C_6 b^{\pp}(Q^2) +{\a_s\o 3\pi}C_8 {m_b\o q^2} c^{\pp}(Q^2)
\left.\right )\left.\right ]\  ,
\end{eqnarray}
\noindent
where
\begin{eqnarray}
a^{\pp}(Q^2)&=&f_\pi (m_B^2-m_{\pi}^2)F_0^{B\pi}(Q^2) \ ,\nonumber\\
b^{\pp}(Q^2)&=&{m_\pi^2\o (m_d+m_u)(m_b-m_u)}a^{\pp}(Q^2) \ ,\nonumber\\
c^{\pp}(Q^2)&=&\left[{{m_B^2-m_\pi^2\o 2(m_b-m_u)}+
       {m_\pi^2\o m_d+m_u}{m_B^2\o m_B^2-m_\pi^2}}\right ] a^{\pp}(Q^2)\ ,
\end{eqnarray}
\noindent with $Q^2=m_\p^2$ and $f_\p=132$MeV.\par
In the calculation of eqs.(15)and (19),
we used the following approximations:
\begin{equation}
 F_0^{B\pi}(m_K^2)\simeq F_0^{B\pi}(m_\pi^2) \simeq F_0^{B\pi}(0)=
F_1^{B\pi}(0) \ ,
\end{equation}
\noindent which  are satisfied within the errors of a few percent
in the pole dominance model of the form factor\cite{BSW}.
In our numerical calculations,
we use the quark mass parameters as
\cite{DATA}
$m_b=4.58$GeV or $5.12$GeV, $m_c=1.45$GeV, $m_s=160$MeV,
 $m_u=5.7$MeV and $m_d=8.7$MeV.
\par
Now we have left with only one unknown parameter $F_0^{B\pi}(0)$
  in the $B\A\kp$ and $B\A\pp$ decay amplitudes except for the
top-quark mass and the CP violating phase in the KM matrix.
 Then, we can predict the ratio of these decay branching ratios
 being independent of the form factor for the fixed phase $\phi$, which
is defined as $V_{ub}=\mid V_{ub}\mid \exp(-i \phi)$.
This ratio is  almost free from the factorization assumption, because
 the ambiguity of this approximation cancels each other in the
numerator and the denominator.
We show the predicted ratio versus $\phi$ in the case of
 $m_b=4.58$GeV and $5.12$GeV with $m_t=140$GeV in fig.1.
Our result  changes only $4\%$ in the region $m_t=120\sim 180$GeV.
However, our result drastically depends on the value of
$\mid V_{ub}/ V_{cb}\mid$ as shown in fig.1, where we use the experimental
 value  $\mid V_{ub}/ V_{cb}\mid=0.08\pm 0.02$\cite{CLEO}
 with $\mid V_{cb}\mid=0.045$.
\begin{center}
 \unitlength=0.7 cm
 \begin{picture}(2.5,2.5)
  \thicklines
  \put(0,0){\framebox(3,1){\bf fig.1}}
 \end{picture}
\end{center}
\par
In order to know the contribution of the penguin process, we give
the ratios of the penguin amplitudes to the tree amplitudes in the
case of $\phi=90^{\circ}$,
$m_b=4.58$GeV and $m_t=140$GeV  for both $B\A\kp$ and $B\A\pp$ processes
as follows:
\begin{equation}
\left | {A({\rm penguin})\o A({\rm tree})}\right | = \Biggl\{\Biggr.
\matrix{4.22\times
              \left ({0.08\o\mid V_{ub}/ V_{cb} \mid} \right )
 \hskip 0.3 cm {\rm for}\  B\A\kp   \cr
       0.22\times \left ({0.08\o\mid V_{ub}/ V_{cb} \mid} \right )
   \hskip 0.3 cm {\rm for}\  B\A\pp } \ \  .
\end{equation}
\par
\noindent  The penguin process dominates
 the $B\A\kp$ decay, but the tree process is not negligible. On the other hand,
 the tree process gives main contributions to
  the $B\A\pp$ decay, although the penguin process is still sizable.\par
Since the experimentally observed branching ratio of $B\A\pp+\kp$ was
given as shown in eq.(1),
we can get the information of the form factor $F_0^{B\p}(0)$
for the fixed $\mid V_{ub}/V_{cb}\mid$ and $\phi$.
We show the branching ratio versus $F_0^{B\pi}(0)$ for
$\phi=90^{\circ}$ and $30^{\circ}$ in fig.2.
The experimental allowed region of the branching ratio
 is the one between the two horizontal dashed-lines in fig.2.
Then, we obtain $F_0^{B\pi}(0)= 0.26\sim 0.55$,
which is consistent with the one in the
BSW model, $F_0^{B\pi}(0)= 0.33$\cite{BSW}.
\begin{center}
 \unitlength=0.7 cm
 \begin{picture}(2.5,2.5)
  \thicklines
  \put(0,0){\framebox(3,1){\bf fig.2}}
 \end{picture}
\end{center}
\par
We comment on the contribution of the charged Higgs boson in the THDM
as a typical new physics candidate.
In contrast with the case of the $B\A X_s\r$ decay,
 the charged Higgs contribution
 cannot provide so sizable enhancement on the $B\A\kp$ decays. This conclusion
is in agreement with the one in the case of $B\A K_X\phi$
 in our previous paper
\cite{NEWHAYASHI}. Our predicted branching ratio
 increases only  in the magnitude of around $10\%$ for
 the case of $m_H=300$\G and $\cot\b=1$, which follow from
the experimental upper bound of the $B\A X_s\r$ decay
\cite{NEWHAYASHI}.\par
 In the above analyses, we have neglected the final state interaction, which
possibly affects the decay widths.
The phase shifts due to the strong final state  interactions
have an effect on the magnitudes of the $B\A\kp$ and $B\A\pp$ decay amplitudes.
First decomposing  their purely weak
(this means "without the final state interaction") amplitudes according to the
 the final state iso-spins and
introducing the corresponding strong phase factor for each iso-spin amplitude,
we obtain the physical decay amplitudes including the final state interaction.
Then, we can readily rewrite,
  the physical decay amplitudes of $B\A K^+\p^-$   in terms of the purely
 weak amplitudes of  $B^0_d\A K^+\p^-$ and $B^0_d\A K^0\p^0$ as
\begin{eqnarray}
 \langle  K^+\pi^-\mid H_{eff}\mid B^0_d\rangle^{phys}& =&
e^{-i\delta_{1/2}}\left\{{1\o 3}(2+e^{i\delta_{\kp}})
\langle  K^+\pi^-\mid H_{eff}\mid B^0_d\rangle^{weak}\right. \nonumber\\
&+&{\sqrt{2}\o 3}(e^{i\delta_{\kp}}-1)
\langle  K^0\pi^0\mid H_{eff}\mid B^0_d\rangle^{weak}\left.\right\}\ ,
\end{eqnarray}
\noindent
where $\delta_{\kp}\equiv \delta_{1/2}-\delta_{3/2}$ and
\begin{eqnarray}
 \langle  \pi^+\pi^-\mid H_{eff}\mid B^0_d\rangle^{phys}& =&
e^{-i\delta_{0}}\left\{
\langle  \pi^+\pi^-\mid H_{eff}\mid B^0_d\rangle^{weak}\right. \nonumber\\
&+&{\sqrt{2}\o 3}(e^{i\delta_{\pp}}-1)
\langle \pi^+\pi^0\mid H_{eff}\mid B^0_d\rangle^{weak}\left.\right\} \ ,
\end{eqnarray}
\noindent
where $\delta_{\pp}\equiv \delta_{0}-\delta_{2}$.
However, the narrow resonances coupled to the  $\kp$ and $\pp$ states
 are not expected at the energy scale of $m_b$. So, the large
 phase shifts are unlikely. Thus, our numerical results
  are not expected to be largely changed by the final state interaction.\par
The summary is given as follows.
We have studied the penguin effect of the $B\A\kp$ and $B\A\pp$ decays
 considering  the recent observed decay branching ratio by CLEO.
We have predicted the ratio of these decay widths, which crucially depends on
the CP violating phase $\phi$ and the still ambiguous KM matrix element
 $\mid V_{ub}\mid$. The experimental information of the $\kp/\pp$ ratio
will serve us these important parameters in the SM model.
 Furthermore, the determination of the form factor will test
many  models of $B$ meson decays.
 We expect that the $K$-$\p$ separation of the $B$ meson decays
will be done in the near future.

\newpage
\topskip 2 cm
\centerline{\large \bf Figure Captions}\par
\vskip 1.5 cm
{\bf figure 1}: \ \ The predicted ratios of
      $BR(B^0_d\A K^+\p^-)/BR(B^0_d\A \p^+\p^-)$ versus
 the phase $\phi$ for $\mid V_{ub}/V_{cb}\mid=0.06, 0.08$ and
 $0.10$.
The solid- and the dashed-lines correspond to the predictions in the case of
 $m_b=4.58$ and  $5.12$GeV with $m_t=140$GeV, respectively.\par
\vskip 1.5 cm
{\bf figure 2}: \ \ The summed  branching ratios of
      $BR(B^0_d\A K^+\p^-)$ and $BR(B^0_d\A\p^+\p^-)$ versus
 the form factor $F_0^{B\pi}(0)$ for $\mid V_{ub}/V_{cb}\mid=0.06,
 0.08$ and $0.10$,
being fixed $m_b=4.58$GeV and $m_t=140$GeV. The solid- and the dashed-lines
 correspond  to the predictions in the case of
 $\phi=90^{\circ}$ and  $30^{\circ}$, respectively.
The upper bound and the lower one of the observed  branching ratio are
  denoted by the two horizontal dashed-lines.
\par
\end{document}